\documentclass[3p,twocolumn]{elsarticle}
\pdfoutput=1
\DeclareGraphicsExtensions{.pdf}

\biboptions{sort&compress}
\usepackage[colorlinks=true,linkcolor=blue, citecolor=blue, urlcolor=blue]{hyperref}
\usepackage{natbib}
\usepackage{color}
\usepackage{multirow}
\usepackage{amssymb}
\usepackage{amsmath}
\usepackage{slashed}
\usepackage{graphicx}
\usepackage{placeins}
\usepackage{array}
\usepackage{dblfloatfix}

\newcommand{\paper}{paper}

\newcommand{\pp}{\mbox{$pp$}}

\newcommand{\jpsi}{\mbox{$J/\psi$}}
\newcommand{\psitwo}{\mbox{$\psi(2S)$}}
\newcommand{\chic}{\mbox{$\chi_c$}}

\newcommand{\ptini}{\mbox{$p_{T,\mathrm{ini}}$}}

\newcommand{\Npart}{\mbox{$N_{\mathrm{part}}$}}

\newcommand{\sqrtsnn}{\mbox{$\sqrt{s_{\mathrm{NN}}}$}}

\newcommand{\Raa}{\mbox{$R_{\rm AA}$}}
\newcommand{\raa}{\mbox{$R_{\rm AA}$}}
\newcommand{\Raaj}{\mbox{$R_{\rm AA}^{J/\psi}$}}
\newcommand{\Raap}{\mbox{$R_{\rm AA}^{\psi(2S)}$}}

\newcommand{\pt}{\mbox{$p_{T}$}}

\newcommand{\cf}            {c_\mathrm{F}}

\journal{Physics Letters B}

\begin{document}

\begin{frontmatter}
\title{
\center{On similarity of jet quenching and charmonia suppression}
}

\author[Charles]{Martin Spousta}
\address[Charles]{Charles University, Faculty of Mathematics and Physics, Prague, Czech Republic}

\begin{abstract}
We quantify the magnitude and the color charge dependence of the medium 
induced parton energy loss in lead-lead collisions at the LHC using the 
data on inclusive jet suppression. The extracted color charge dependence 
shows that the difference between the in-medium loss of quarks and 
gluons is consistent with the difference between the radiation of quarks 
and gluons in the vacuum. Then, we examine the energy loss of prompt 
charmonia and we point to a remarkable similarity between the quenching 
of light-quark-initiated jets and the prompt charmonia suppression. 
Finally, we discuss possible sources of this similarity.

\end{abstract}

\end{frontmatter}

\newlength{\fighalfwidth}
\setlength{\fighalfwidth}{0.49\textwidth}

\section{Introduction}
\label{sec:intro}

  Collisions between lead nuclei at the LHC produce colored medium where 
relevant degrees of freedom are deconfined 
quarks and gluons \cite{Karsch:2003jg}.
  Hard scattering interactions in perturbative quantum chromodynamics 
(pQCD) lead to production of two highly virtual back-to-back partons (in 
the leading order of pQCD) which subsequently evolve as parton showers, 
hadronize, and are experimentally observed as back-to-back dijet. 
If partons traverse on their path a colored medium they lose energy 
which can be seen as modification of jet yields and jet properties. 
This phenomenon is commonly called ``jet quenching'' 
\cite{Wiedemann:2009sh,Majumder:2010qh,Mehtar-Tani:2013pia}. 
  The first evidence of jet quenching at the LHC was provided by the 
measurement of jet pairs \cite{Aad:2010bu,Chatrchyan:2011sx}. Often, the 
magnitude of jet quenching is quantified by the nuclear modification 
factor, \Raa ,
  which is the ratio of per-event-normalized yields in lead-lead 
(Pb+Pb) collisions to a cross-section in proton-proton ($pp$) 
collisions scaled by the nuclear overlap function.
  A precise measurement of jet \Raa\ was recently 
provided in Ref.~\cite{Aad:2014bxa}. 
  Understanding the jet quenching means understanding the interaction of 
energetic color charges with the deconfined medium and the properties of this 
medium.

  Not only jets but also charmonia can be used as tools to reveal the 
properties of the medium and its interactions.
  The production of charmonia in elementary collisions is often 
described in a nonrelativistic QCD effective-field-theory (EFT) framework 
\cite{Bodwin:1994jh}. In that theory, first, the $c\bar{c}$ pair is produced either 
in a color singlet or color octet state. This ``pre-resonant'' $c 
\bar{c}$ pair then binds into a physical charmonium by non-perturbative 
evolution described in terms of long-distance matrix elements. The 
pre-resonant pair in the color octet state changes its color and spin
  by radiating off gluon(s) when evolving to the 
physical quarkonium state while the pair in the color singlet state 
retains these quantum numbers unchanged. 

  In nucleus-nucleus collisions, the production of charmonia was 
observed to be strongly suppressed with respect to proton-proton yield 
at LHC, RHIC, and SPS \cite{Andronic:2015wma}. Generally, charmonia 
suppression can be described using the EFT framework at finite 
temperature in the limit of weakly coupled medium or in the lattice-QCD 
in the strongly coupled regime \cite{Brambilla:2004wf}. In the EFT 
 framework at finite temperature
 the thermal part of $c\bar{c}$ potential has both a real and imaginary part \cite{Brambilla:2008cx}. 
The real part of the $c\bar{c}$ potential is connected with color 
screening of the potential in the deconfined medium. The imaginary part 
of the potential is connected with thermal decay and it is argued that 
it may be a dominant mechanism for the charmonia suppression within the 
EFT framework \cite{Brambilla:2010cs}. Since heavy quarks are not fully 
thermalized in the medium, further model assumptions need to be made in 
order to predict charmonium production rates. Examples of these models 
are transport models, statistical hadronization models, collisional 
dissociation models, or comover models which are reviewed in 
Ref.~\cite{Andronic:2015wma}. Different mechanisms leading to charmonia 
suppression that are implemented in these models play different role at 
different kinematic regions and center-of-mass energies which makes the 
interpretation of similarities or differences in the charmonia 
suppression between the LHC, RHIC, and SPS complicated.

  In proton-nucleus (or deuteron-nucleus) collisions, the production of 
charmonia was observed to be also modified with respect to proton-proton 
yield although the modifications are generally smaller than those seen 
in nucleus-nucleus collisions \cite{Andronic:2015wma}. 
  These modifications are due to an interplay of various affects such as 
effects of nuclear modifications to parton distribution functions 
(shadowing, anti-shadowing, EMC effect), multiple parton scattering, or 
nuclear absorption of bound state. These ``cold nuclear matter effects'' 
(CNM effects) are present in nucleus-nucleus collisions as well with a 
strength that depends on the choice of kinematic region. Therefore, they 
need to be considered when interpreting the nucleus-nucleus data.

  This \paper\ elaborates on the quantification of the jet quenching 
and it explores a similarity in the suppression of jets and prompt 
charmonia (that is those not coming from weak decay of $B$-hadrons)  
in $\sqrtsnn=2.76$~TeV Pb+Pb collisions at the LHC. The work 
is organized as follows. First, the effective suppression parameters of 
high-\pt\ quarks and gluons are extracted from jet suppression 
measurements. Then, these parameters are used to model the suppression 
of prompt charmonia. The result is compared with several recent 
measurements. Finally, the connection between the energy loss of jets 
and the charmonia suppression is discussed.

\section{Single jet suppression}
\label{sec:jets}

It was shown in a recent paper on the interpretation of single jet 
suppression measurements at the LHC \cite{Spousta:2015fca} that many of 
the features seen in the data are driven by a primordial parton spectrum 
and the different quark and gluon energy loss.
  Further, it is concluded in that study that the suppression data on 
jets can be described by an effective quenching model (hereafter called EQ model) in 
which partons lose energy depending on their initial transverse 
momentum, \ptini , and an effective color factor, $\cf$, such that the 
total energy lost by a parton is 
  \begin{equation}
  \label{eqn1}
  \Delta \pt\ = \cf \cdot s \cdot \bigg( \frac{ \ptini }{ p_{T,0} } \bigg)^\alpha.
  \end{equation}
  Here $s$ and $\alpha$ are free parameters of the model, \ptini\ is the 
transverse momentum of a parton initiating a jet and $p_{T,0}$ is an 
arbitrary scale, here and in Ref.~\cite{Spousta:2015fca} set to 40~GeV.
  The parameter $\cf$ was fixed to be 1 and 9/4 for light-quark-initiated 
jets and gluon-initiated jets, respectively. Input to the model was the 
parameterization of the unquenched jet-\pt\ spectra which was obtained 
from the PYTHIA8 generator \cite{Sjostrand:2014zea}.
  The EQ model successfully described the representative data on the 
single jet suppression at high-\pt : the inclusive jet \Raa\ in various 
rapidity intervals \cite{Aad:2014bxa}, trends in the jet fragmentation 
functions \cite{Chatrchyan:2014ava,Aad:2014wha} at $\pt \gtrsim 10$~GeV 
and the inclusive charged particle \Raa\ at $\pt \gtrsim 20$~GeV 
\cite{CMS:2012aa}.

  The EQ model represents a model with minimal assumptions on the 
physics of the jet quenching process. The dynamics of the quenching 
process as well as the properties of the medium are encapsulated in a 
few free parameters.
  The fact that not only the inclusive jet suppression but also the jet 
structure is largely captured by the model speaks in favor of 
a physics picture in which a significant part of a parton shower remains 
unresolved by the medium. That may happen if the parton shower loses its 
energy coherently and subsequently fragments as in the vacuum. Indeed, 
it was recognized in several theoretical papers that such color 
coherence effects play an important role in the jet quenching process 
\cite{MehtarTani:2010ma,MehtarTani:2011tz,CasalderreySolana:2012ef,Blaizot:2013hx}.
  If the parton shower, or its large part, radiates as one object, one 
can ask if it is possible to find some similarities between the 
suppression of jets and a suppression of other objects with an 
internal structure.
  One such candidate are the charmonia.
Before turning the attention to that question, 
the original model calls for an extension:
  the color factor should be extracted from the data and a possible 
dependence on the input spectra should be quantified.

  The choice of $\cf = C_A/C_F = 9/4$ used in the EQ model for gluons 
represents an assumption on the difference in the probability to radiate 
a gluon from a gluon and quark source in the vacuum in the 
$Q \rightarrow \infty$ limit \cite{Konishi:1979cb} or the soft limit 
\cite{Khoze:1997zq}. 
  The value of $\cf$ can be different due to both the neglected 
non-leading corrections \cite{Capella:1999ms} and the presence of the deconfined medium.
  Thus, a global fit has been performed to extract $s$, $\alpha$ and $\cf$ 
simultaneously. This has been done by minimizing the difference between the 
EQ model and measured jet \Raa\ \cite{Aad:2014bxa} in different rapidity 
intervals and centrality bins.
  This procedure follows the logic of extracting this factor in the vacuum 
\cite{Acosta:2004js,Abbiendi:1999pi}.

  To test for a sensitivity of the extracted suppression parameters on 
the input jet spectra, the POWHEG MC generator 
\cite{Alioli:2010xd,Alioli:2010xa} interfaced to PYTHIA8 was used as an 
alternative to PYTHIA8 to provide a full next-to-leading order (NLO) 
jet \pt\ spectra. 
  This NLO simulation was repeated three times using different input PDF 
sets (CT10 \cite{Lai:2010vv}, MSTW2008NLO \cite{Martin:2009iq}, and 
NNPDF2.0 \cite{Ball:2010de}) which were selected to correspond to those 
used in the precise measurement of the jet cross-section in $pp$ 
collisions at $\sqrt{s} = 7$~TeV \cite{Chatrchyan:2011ab}. These three 
sets of jet-\pt\ spectra were then parameterized and used along with the 
nominal PYTHIA8 as an input to the model which allowed to quantify 
uncertainties on extracted parameters.

\begin{table}
\begin{center}
\begin{tabular}{|l|c|} \hline
\multirow{2}{*}{$s = x \cdot \Npart + y$}       & $x = (12.3 \pm 1.4) \cdot 10^{-3}$~GeV,   \\
                                                & $y = 1.5 \pm 0.2$~GeV \\ \hline
$\alpha$                                        & $0.52 \pm 0.02$     \\ \hline
$\cf$                                           & $1.78 \pm 0.12$     \\ \hline
\end{tabular}
\end{center}
\caption{
  Parameters of the effective quenching model extracted from data.
  The $\chi^2/n_{dof}=0.63$ for $n_{dof}=287$. 
  For details see the text.
}
\label{tab:tab1}
\end{table}

\begin{figure*}[h]
\begin{center}
\includegraphics[width=0.32\textwidth]{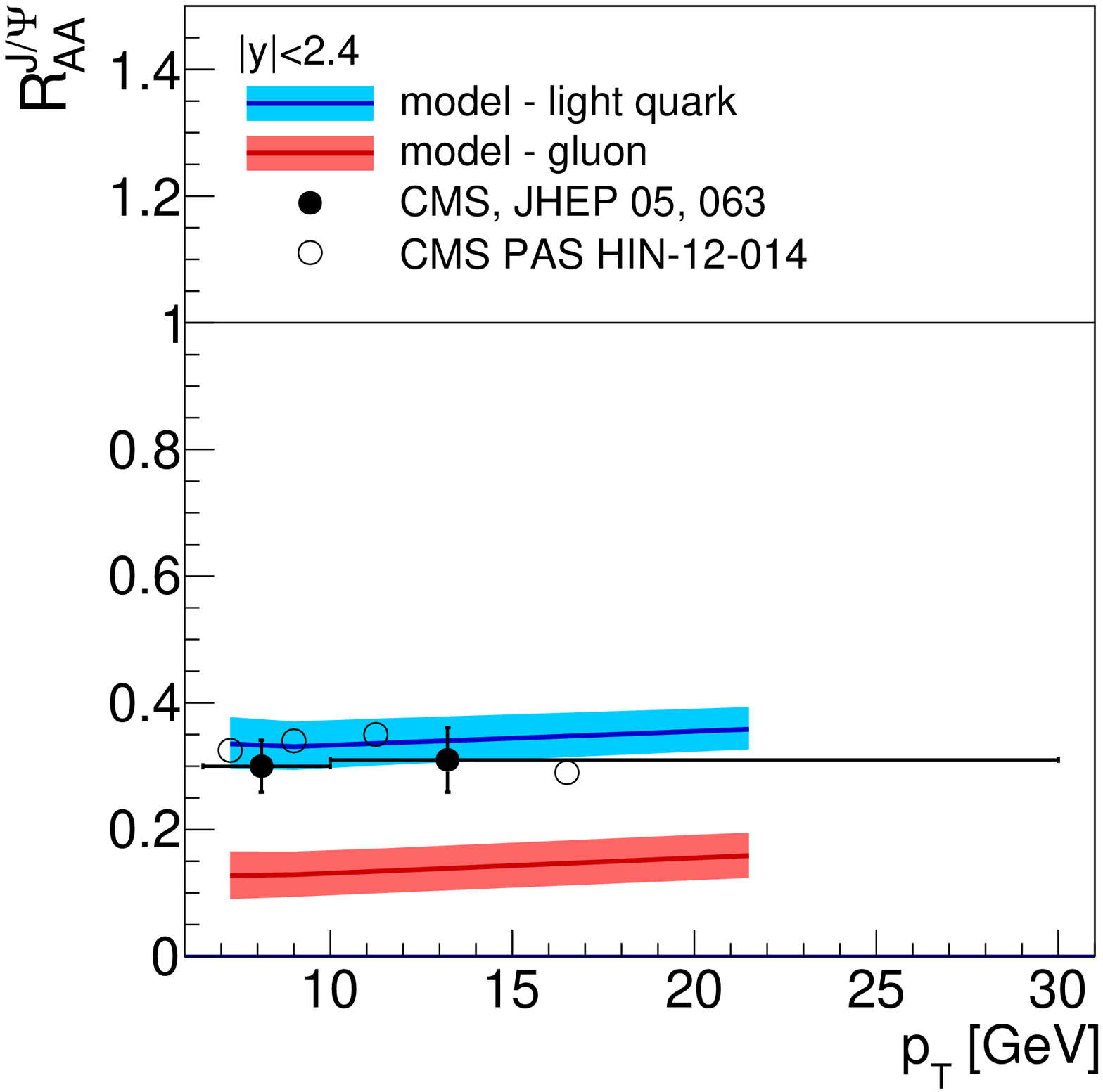}
\includegraphics[width=0.32\textwidth]{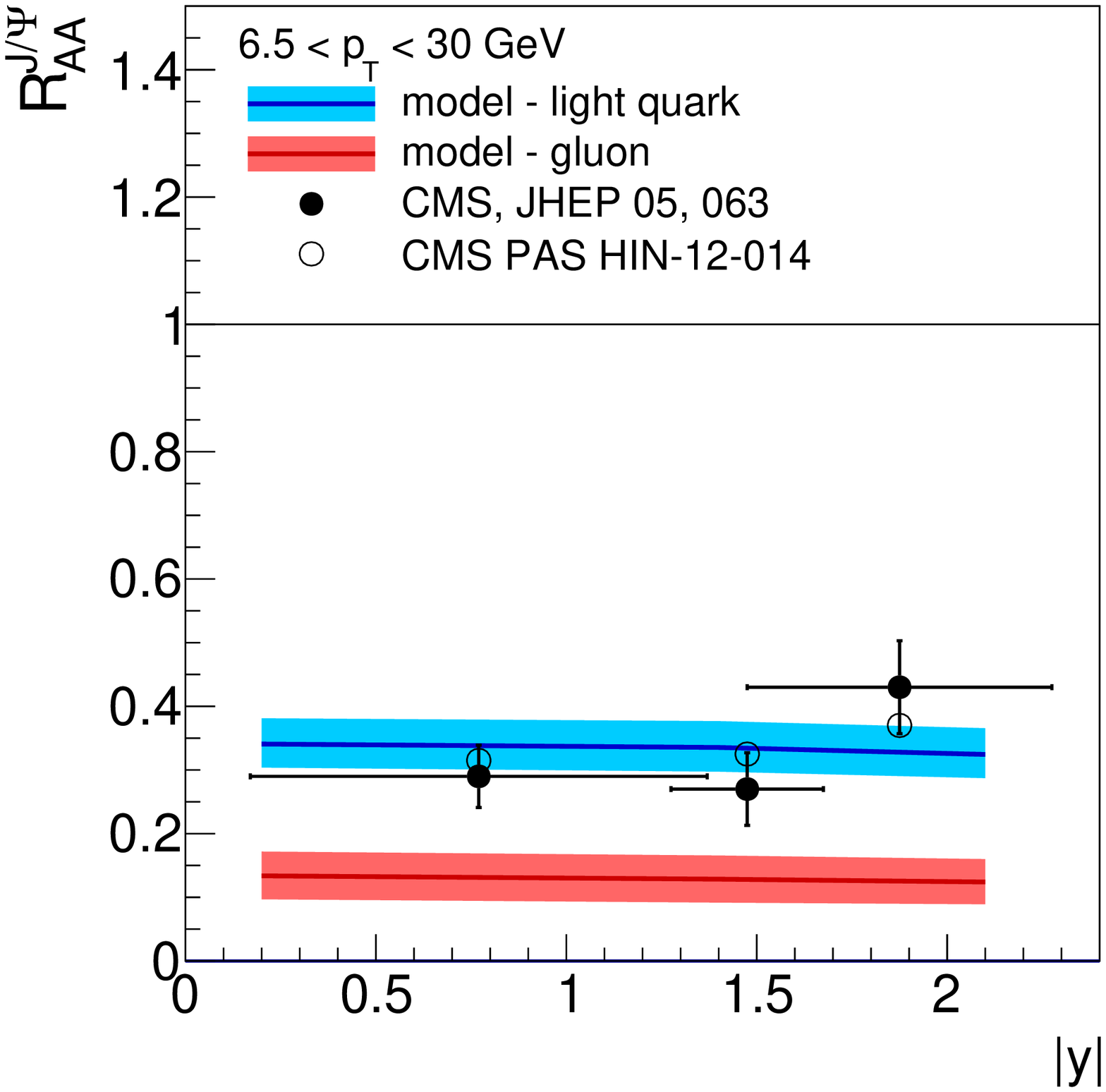}
\includegraphics[width=0.32\textwidth]{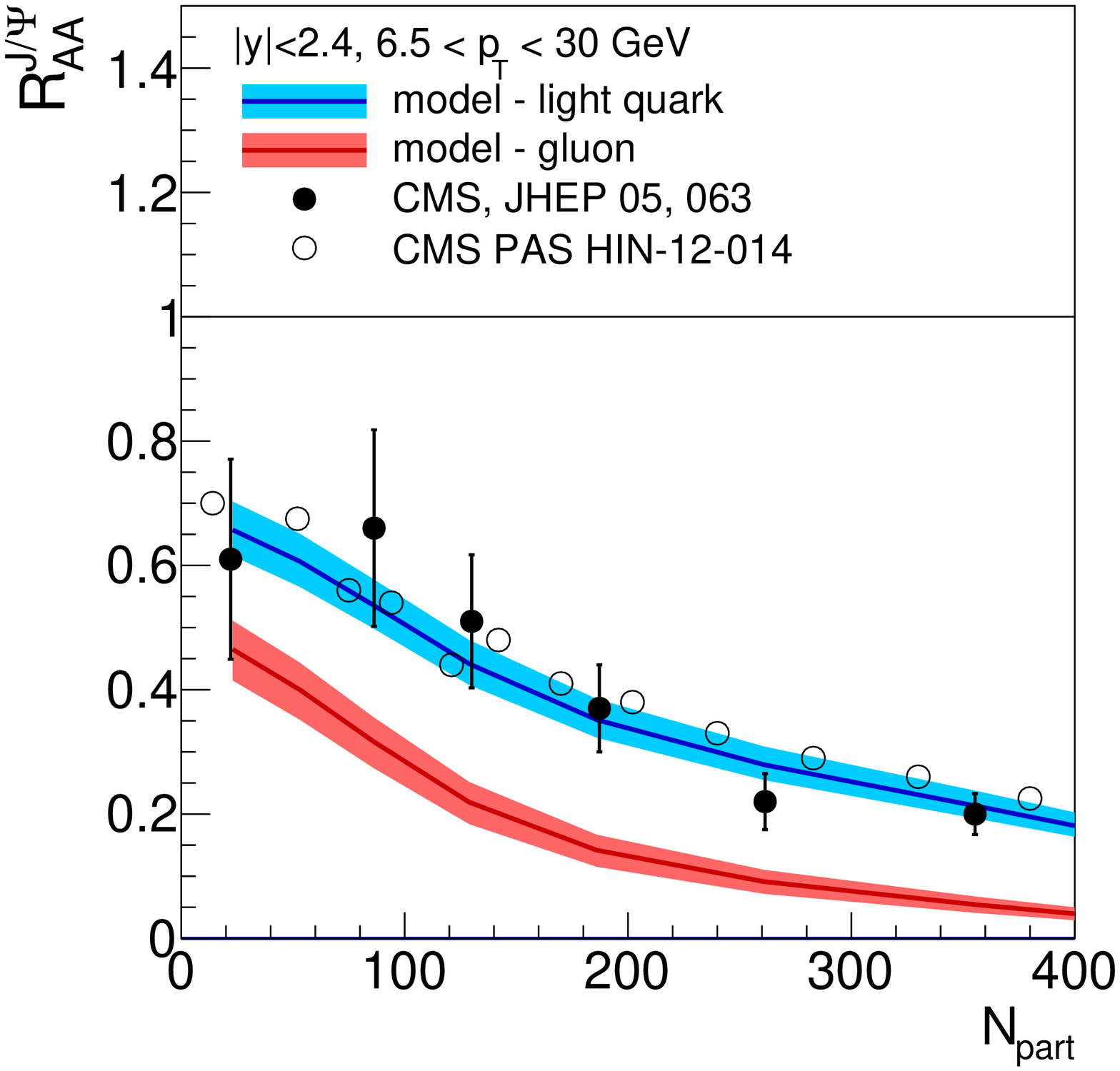}
\end{center}
\caption{
  The \Raaj\ predicted by the model compared to the data measured 
differentially in \pt\ of \jpsi\ {\it (left)}, rapidity of \jpsi\ {\it 
(middle)}, and the number of participants {\it (right)} 
\cite{Chatrchyan:2012np,CMS:2012vxa}. 
  The model is evaluated in two 
configurations of the effective color factor -- the color factor 
corresponding to light quarks and gluons.
  }
\label{fig:fig1}
\end{figure*}

  The values of parameters from the global fit are summarized in 
Tab.~\ref{tab:tab1}.
  The global fit has revealed three interesting properties related to 
the energy loss: 1) the magnitude of the energy loss, $s$, depends 
linearly on the $\Npart$ (thus $s$ is parameterized as a function of 
$\Npart$ in Tab.~\ref{tab:tab1}); 2) the power $\alpha$ is constant as a 
function of $\Npart$ (already observed
in Ref.~\cite{Spousta:2015fca}); 3) $\cf$ is consistent with the value 
calculated and measured in the vacuum which is $\approx 1.8$ at the jet 
hardness $Q=100$~GeV \cite{Capella:1999ms,Acosta:2004js}.
  This represents a precise effective 
quantification of the jet quenching in $\sqrtsnn = 2.76$~TeV Pb+Pb data.
  Extracting $\cf$ from the data using a model with minimal assumptions 
on the physics of the jet quenching should help to constrain the theory 
of this phenomenon.

\section{Quarkonia suppression}
\label{sec:charmonia}

There is no unique interpretation of the charmonia suppression 
measurements as of now \cite{Andronic:2015wma}. New measurements at the 
LHC \cite{Aad:2010aa,Chatrchyan:2012np,Abelev:2013ila,Khachatryan:2014bva,Adam:2015rba,CMS:2012vxa}
  should provide more insight to the mechanism of 
the charmonia suppression. The new precise measurements of the prompt \jpsi\ in 
the muon channel \cite{Chatrchyan:2012np,CMS:2012vxa} showed that the nuclear modification 
factor, \Raaj , reaches a value of $\sim 0.2$ in 
the most central collisions ($\Npart \gtrsim 350$), continuously grows up to a 
value of $\sim 0.6-0.7$ reached in the most peripheral collisions 
($\Npart \lesssim 50$). The \Raaj\ exhibits only a weak (if any) dependence on 
the \jpsi\ momentum in the region of $\pt = 6.5 - 30$~GeV and $|y|<2.4$. 
The dependence of \Raaj\ on the rapidity is also weak. More recently,
a prompt production of \psitwo\ was also measured
in terms of a 
double ratio of measured yields, $(N_{\psi(2S)} / N_{J/\psi})|_\mathrm{Pb+Pb} / 
(N_{\psi(2S)} / N_{J/\psi})|_{pp} = \Raap / \Raaj$ \cite{Khachatryan:2014bva}.
It was shown that \psitwo\ yields are suppressed 
by a factor of $\sim 2$ with respect to \jpsi\ in the range $|y|<1.6$ 
and $6.5 < \pt < 30$~GeV.

\begin{figure}
\begin{center}
\includegraphics[width=0.32\textwidth]{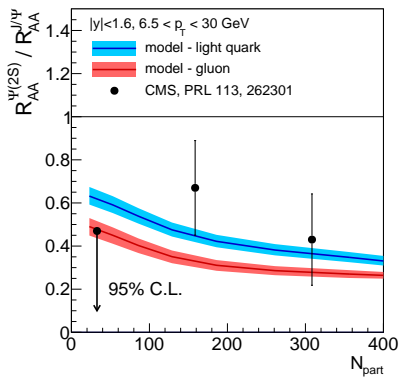}
\end{center}
\caption{
  The ratio of \Raap\ to \Raaj\ predicted by the model compared to the 
data \cite{Khachatryan:2014bva}. 
  The model is evaluated in two 
configurations of the effective color factor -- the color factor 
corresponding to light quarks and gluons.
  }
\label{fig:fig2}
\end{figure}

To test the idea of similarity in the physics of jet quenching and 
prompt charmonia suppression the EQ model described in the previous section has been
employed. The input to the model are \pt\ spectra of \jpsi\ and \psitwo\ 
and effective parameters obtained from the analysis of jet \Raa . The 
\pt\ spectra were obtained from PYTHIA8 (the same initial setup as in 
Ref.~\cite{Spousta:2015fca} was used) which was reweighted to reproduce the 
data measured in $pp$ collisions at the 2.76 TeV \cite{ATLAS:2015pua}.
  It was found that PYTHIA reproduces well the shape of the \jpsi\ \pt\ 
spectra even without reweighting, while a reweighting was needed 
to reproduce the \pt\ spectra of \psitwo (weights range from 0.2 to 6).
  The realistic $\pt$ spectra were then used as an input to the EQ model 
which was run with two different settings of the color factor: first, 
corresponding to the color factor for the energy loss of light-quark 
initiated jets (defined to be one), and second, corresponding to the 
color factor extracted for the energy loss of gluon-initiated jets.
  The comparison of the data with the model is shown in 
Fig.~\ref{fig:fig1}. 
  An excellent agreement of the model with the data for the case of the 
light-quark energy loss is seen. The \Npart\ dependence of \Raaj , its 
\pt\ and rapidity dependence are reproduced. 

  The ability of the model to reproduce the suppression of \psitwo\ is 
shown in the Fig.~\ref{fig:fig2} where the model is compared with the 
data on the ratio of nuclear modification factors, $\Raap / \Raaj$, from 
Ref.~\cite{Khachatryan:2014bva}. Remarkably, a good agreement between 
the model and the data is seen. The model reproduces the measurement 
well except for the most peripheral collisions. Here, however,
no significant 
\psitwo\ signal was measured by CMS and consequently only a limit at 
95\% confidence level on the value of the ratio was provided.

\begin{figure}[t]
\begin{center}
\includegraphics[width=0.32\textwidth]{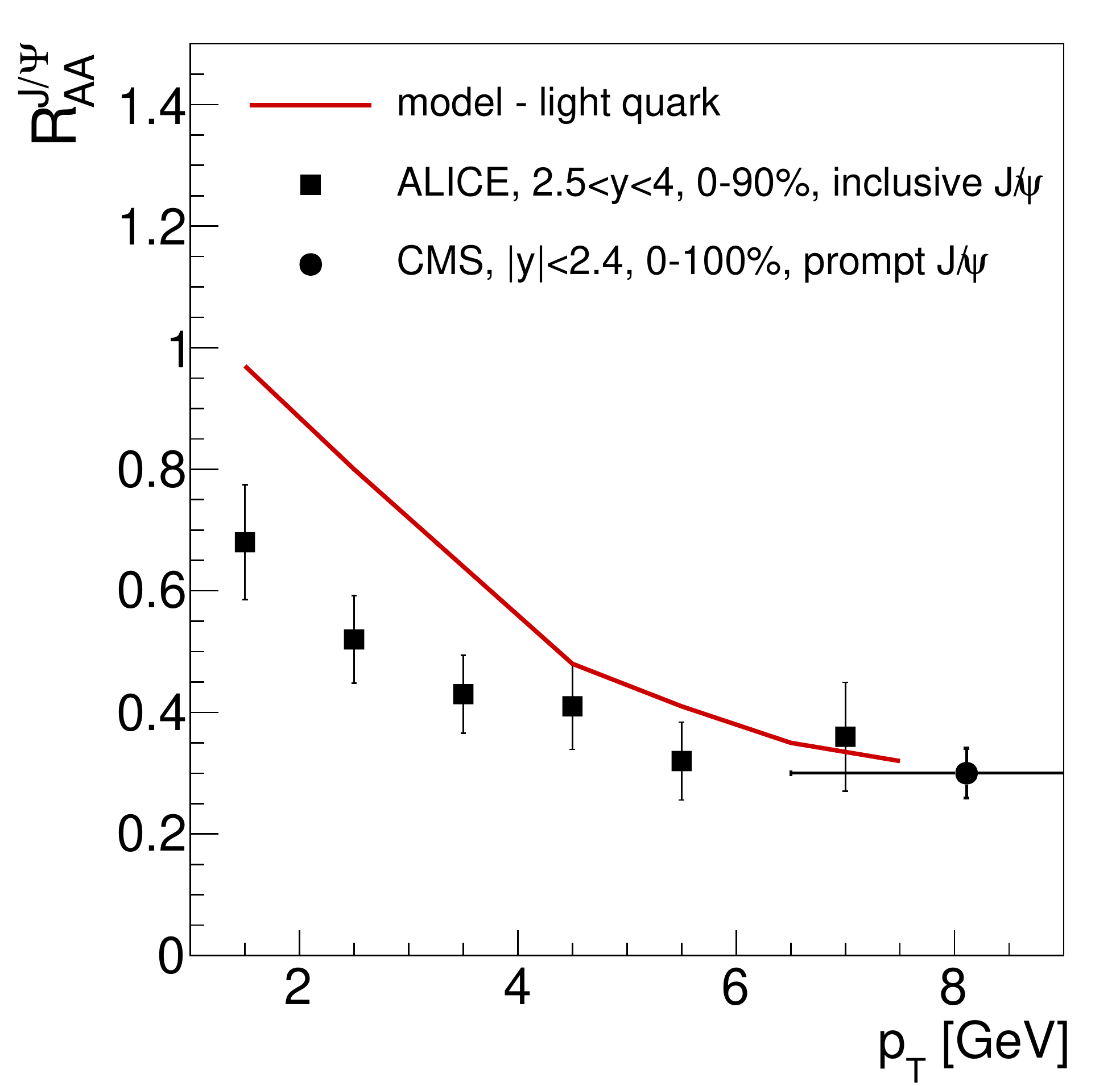}
\end{center}
\caption{
  The \Raaj\ predicted by the model compared to the data measured by 
ALICE at low \pt\ in the forward rapidity region \cite{Abelev:2012rv}.
  }
\label{fig:fig3}
\end{figure}

  The striking similarity between the measured \jpsi\ and \psitwo\
suppression and the energy loss of jets suggests that the radiative energy
loss may be a dominant contribution to the energy loss of charmonia in the
studied kinematic region as we will discuss in the next section.
  Now, we will discuss impact of feed-down contributions from excited 
states on presented results and the role of CNM effects.
  The measured charmonia production generally includes direct production 
from the hard interaction, as well as charmonium feed-down from excited 
states.
  While the feed-down contribution from excited resonances to \psitwo\ 
state is minimal, the feed-down contribution to \jpsi\ is significant 
\cite{Andronic:2015wma}. Thus, it is important to evaluate the impact of 
feed-down on presented results.
  The size of feed-down contribution from \psitwo\ and \chic\ decays to
\jpsi\ were estimated to be 8\% and 25\%, respectively
\cite{Faccioli:2008ir}.
  The \pt\ dependence of the dominant feed-down contribution of \chic\
to \jpsi\ was then evaluated in \pp\ collisions at the LHC
\cite{LHCb:2012af,ATLAS:2014ala}. In particular, for the central 
rapidity region of $|y|<0.75$, it was found that the contribution from
\chic\ is on average 25\% varying by only 2\% in the \pt\ interval of
$10-30$~GeV.
  Also measured in \pp\ collisions was the \pt\ spectrum of \chic\ 
\cite{ATLAS:2014ala}.
  To access the impact of the feed-down on presented results, the 
measured \chic\ and \psitwo\ spectra and fractions of prompt \jpsi\ 
produced in \chic\ and \psitwo\ decays, respectively, were used to 
reweight the initial spectrum of \jpsi\ such that its shape reflects a 
contribution from excited states. The \raa\ of \jpsi\ was then 
recalculated using this new spectrum. The new reweighted spectrum 
characterizes a physics picture in which it is always the excited state 
which loses the energy in the medium while the \jpsi\ resulting from the 
decay of that excited state does not lose the energy. More realistic 
scenario is that only a fraction of \chic\ or \psitwo\ loses the energy 
while the rest first decays and then it loses the energy as \jpsi . 
Since we do not know such a fraction, the difference of reweighted \raa\ 
and original \raa\ was used as an estimate of the uncertainty of 
presented results. Maximal difference between the reweighted and 
original \raa\ was found to be less then 10\%.
  This estimate is similar to the estimate of the impact of feed-down on 
the nuclear modification factor measured in d+Au collisions at RHIC 
which was found to be 5\% \cite{Adare:2013ezl}. The uncertainty in the 
determination of \Raa\ due to the feed-down effects was combined with 
the uncertainty in the determination of \Raa\ due to uncertainties in 
parameters of the EQ model and is plotted on 
Figures~\ref{fig:fig1} and \ref{fig:fig2} as a shaded band.

  Besides feed-down effects, the interpretation of presented results 
might be obscured by initial-state effects or effects from final-state 
nuclear interactions that are unrelated to the presence of the hot 
deconfined medium, i.e. by CNM effects. Sizable suppression of \jpsi\ 
was seen in proton-lead (p+Pb) collisions in the forward rapidity region 
or at low \pt\ ($\pt \lesssim 6$~GeV) at the LHC 
\cite{Adam:2015iga,Abelev:2013yxa,Aaij:2013zxa} which clearly suggests 
that these effects come into the play. On the contrary, at high \pt\ 
($\pt \gtrsim 6$~GeV) in the midrapidity region at the LHC, the nuclear 
modification factor measured in p+Pb collisions is consistent with unity 
within the precision of measurements \cite{Adam:2015iga,ATLAS:2015pua}. 
This observation is consistent with the estimate done using EKS98 nPDFs 
\cite{Eskola:1998df} which quantifies the impact of CNM effects, namely 
the shadowing, on the \Raa\ to be less than 10\% in the kinematic region 
probed in this \paper\ \cite{Andronic:2015wma}.
  While this suggests that CNM effects should not affect the 
interpretation of the similarity of the jet quenching and charmonia 
suppression, there might still be a group of CNM effects that is common 
for both, charmonia and jets. Suggestive of that is a sizable 
suppression seen in the most peripheral bin for both, the charmonia and 
jets.

  Figures \ref{fig:fig1} and \ref{fig:fig2} represent a selection of all available 
LHC data on prompt \jpsi\ and \psitwo\ for which corresponding \pp\ 
reference spectra exist which allows an exact data-to-model comparison. 
Besides those high-\pt\ \raa\ measurements, ALICE Collaboration provided 
also measurement of \Raaj\ \cite{Abelev:2013ila} for which 
the corresponding \pp\ spectra were measured \cite{Abelev:2012kr}. This 
measurement presents \raa\ of inclusive \jpsi\ (that is with no 
separation of non-prompt \jpsi\ contribution which is however smaller 
than 15\% \cite{Book:2014mia}) in the forward rapidity region and at low 
\pt . As discussed before, in this kinematic region CNM effects 
play a significant role. It was also shown in past that recombination 
and statistical hadronization processes likely play a significant role in this 
kinematic region 
\cite{Zhao:2011cv,Liu:2009nb,Ferreiro:2012rq,Andronic:2011yq}. Thus, we 
cannot expect that the model presented here will reproduce the data. 
Nevertheless, it is instructive to quantify a departure of the model 
from the data. This is shown in Fig.~\ref{fig:fig3}. One can see that 
the model can reproduce the \Raaj\ at $\pt \gtrsim 6$ GeV where ALICE 
data in the forward region agrees with CMS data measured in the 
midrapidity region. On the contrary, a significant departure 
of the model from the data is seen at low \pt .

\section{Discussion}

  The ability of the EQ model to reproduce the suppression data on 
\jpsi\ and \psitwo\ suggests that the measured shape of the suppression 
at high-\pt\ is driven by the shape of the initial spectra of charmonia 
and that the radiative energy loss may play a dominant role at high-\pt 
.
  We will leave the full analysis of this observation and the 
observation of similarity between the light-quark suppression and the 
charmonia suppression for separate works. Here, we will only 
summarize straightforward leading order calculations searching for a 
similarity between the radiation of quarks and charmonium. In 
particular, we discuss two basic scenarios: 1) charmonium is produced 
early in the collision in the color octet state and radiates coherently; 
2) charmonium is produced late, from quarks or gluons that are already 
quenched.

  For the first scenario, ratio is calculated of the probability of 
radiating a gluon from a charmonium in the color octet state, $P_\psi$, 
to the probability of radiating a gluon from a single light quark, $P_q$. This 
ratio is calculated in the collinear limit, leading to no change in the 
spin of charmonium due to the radiation, and under the assumption that 
the $c\bar{c}$ pair is not dissociated and radiates coherently in the range 
of virtualities from $Q_{max}^\psi$ to $Q_{min}^\psi$. The result is
  \begin{equation}
  \label{eqn:eqn2}
  \frac{P_\psi
  (Q_{min}^{\psi},Q_{max}^{\psi})}
       {P_q
  (Q_{min},Q_{max})} =
  \frac{c_\psi}{C_F} \bigg( 1 + \ln \frac{k_{max}}{k_{min}} \cdot \ln^{-1}\frac{Q_{max}}{Q_{min}} \bigg),
  \end{equation}
  where $c_\psi=C_A$ is the color factor for the radiation from 
charmonium. Constant $k$ relates virtualities in the two processes, 
$Q^\psi = kQ$ \footnote{Equation (\ref{eqn:eqn2}) holds also for the 
case of massive splitting functions calculated in the quasi-collinear 
limit \cite{Catani:2000ef}.}. The ratio (\ref{eqn:eqn2}) is equal to one 
if the virtuality range differs between the two processes (e.g. 
$k_{max}/k_{min} \approx 1/9$ for $Q_{min}=0.2$~GeV and 
$Q_{max}=10$~GeV). If this is the case, Eq.~(\ref{eqn:eqn2}) predicts 
that the \Raa\ of \jpsi\ should start to deviate from the \Raa\ of 
light-quark initiated jets at higher transverse momenta due to the slow 
logarithmic dependence on the virtuality. In particular, at $\pt = 
50$~GeV, the \Raa\ of light-quark initiated jets should be 1.3 times 
larger compared to the measured \Raa\ of \jpsi . This prediction can be 
tested in the LHC run 2 data.
  Not quantified in the formula is the dead cone effect which may also 
decrease the resulting radiation of the charmonium 
\cite{Dokshitzer:2001zm,Armesto:2003jh}.
  Another factor not accounted for in this scenario is a presence of the 
charmonium in the color singlet state which might however represent 
a small contribution compared to the color octet state 
\cite{Baranov:2015laa}.
  Irrespective of these important details, Eq.~(\ref{eqn:eqn2}) implies 
that a simple expectation that the radiative energy loss of charmonia 
should be similar to the radiative energy loss of gluon-initiated jets 
is not valid since not only the color factors enter here but also 
kinematic range over which the two systems radiate need to be 
considered.

In the second scenario, the charmonium is produced late, from quarks or 
gluons that are already quenched.  While the correlation between the 
\pt\ of the initial parton and the \pt\ of charmonium is rather strong, 
less steep spectra of charmonia would lead to generally larger \Raa\ of 
charmonia than the \Raa\ of partons. This was explicitly checked by 
modeling using PYTHIA. In this scenario, the similarity between the 
suppression of charmonia and light quarks would be accidental, steaming 
from an intriguing combination of quark and gluon energy loss which 
disfavors this scenario.

\section{Summary}

  This \paper\ has presented a quantification of the jet quenching in 
$\sqrtsnn = 2.76$~TeV lead-lead collisions which is based on an extension of 
the effective quenching model \cite{Spousta:2015fca}.
  In particular, extracted from the data was the effective color 
factor, $\cf$, characterizing a difference in the probability to radiate 
a gluon from a gluon and quark source. The extracted value, $\cf = 1.78 
\pm 0.12$, is consistent with the value obtained for vacuum \cite{Capella:1999ms,Acosta:2004js}.
  Using the quantification of the energy loss of jets, a similarity 
between the suppression of jets and prompt charmonia was explored. A 
striking similarity between the measured \jpsi\ and \psitwo\ suppression 
and the energy loss of light-quark initiated jets was seen. 
  While this observation requires a thorough theoretical analysis, it 
suggests that the radiative energy loss may be a dominant contribution to 
the charmonia suppression at $\pt \gtrsim 6$~GeV at the LHC.
  Quantification of the magnitude of the jet quenching together with 
observations made in this \paper\ should improve the understanding of 
physics mechanism behind both, the jet quenching and charmonia 
suppression. They may also contribute to better understanding of 
general aspects of charmonia formation.

\section*{Acknowledgment}
 I'm grateful to Alexander Kup\v co, Konrad Tywoniuk, 
 Brian Cole, Ji\v r\' i 
 Dolej\v s\' i, Daniel Scheirich, Yen-Jie Lee and 
 Camelia Mironov for useful discussions.
   This work was supported by Charles University Research Development 
Scheme (PRVOUK) P45, UNCE 204020/2012 and MSMT grant INGO II LG15052.

\bibliography{onia,heavyIonsExperiment,heavyIonsTheory,strongIntExperiment,strongIntTheory}

\begin{thebibliography}{10}
\expandafter\ifx\csname url\endcsname\relax
  \def\url#1{\texttt{#1}}\fi
\expandafter\ifx\csname urlprefix\endcsname\relax\def\urlprefix{URL }\fi
\expandafter\ifx\csname href\endcsname\relax
  \def\href#1#2{#2} \def\path#1{#1}\fi

\bibitem{Karsch:2003jg}
F.~Karsch, E.~Laermann, {Thermodynamics and in medium hadron properties from
  lattice QCD}\href {http://arxiv.org/abs/hep-lat/0305025}
  {\path{arXiv:hep-lat/0305025}}.

\bibitem{Wiedemann:2009sh}
U.~A. Wiedemann, {Jet Quenching in Heavy Ion Collisions} (2010)
  521--562[Landolt-Bornstein23,521(2010)].
\newblock \href {http://arxiv.org/abs/0908.2306} {\path{arXiv:0908.2306}},
  \href {http://dx.doi.org/10.1007/978-3-642-01539-7_17}
  {\path{doi:10.1007/978-3-642-01539-7_17}}.

\bibitem{Majumder:2010qh}
A.~Majumder, M.~Van~Leeuwen, {The Theory and Phenomenology of Perturbative QCD
  Based Jet Quenching}, Prog. Part. Nucl. Phys. A66 (2011) 41--92.
\newblock \href {http://arxiv.org/abs/1002.2206} {\path{arXiv:1002.2206}},
  \href {http://dx.doi.org/10.1016/j.ppnp.2010.09.001}
  {\path{doi:10.1016/j.ppnp.2010.09.001}}.

\bibitem{Mehtar-Tani:2013pia}
Y.~Mehtar-Tani, J.~G. Milhano, K.~Tywoniuk, {Jet physics in heavy-ion
  collisions}, Int. J. Mod. Phys. A28 (2013) 1340013.
\newblock \href {http://arxiv.org/abs/1302.2579} {\path{arXiv:1302.2579}},
  \href {http://dx.doi.org/10.1142/S0217751X13400137}
  {\path{doi:10.1142/S0217751X13400137}}.

\bibitem{Aad:2010bu}
{ATLAS Collaboration}, {Observation of a Centrality-Dependent Dijet Asymmetry
  in Lead-Lead Collisions at $\sqrt{s_{NN}}=2.77$ TeV with the ATLAS Detector
  at the LHC}, Phys.~Rev.~Lett. 105 (2010) 252303.
\newblock \href {http://arxiv.org/abs/1011.6182} {\path{arXiv:1011.6182}},
  \href {http://dx.doi.org/10.1103/PhysRevLett.105.252303}
  {\path{doi:10.1103/PhysRevLett.105.252303}}.

\bibitem{Chatrchyan:2011sx}
{CMS Collaboration}, {Observation and studies of jet quenching in PbPb
  collisions at nucleon-nucleon center-of-mass energy = 2.76 TeV}, Phys. Rev.
  C84 (2011) 024906.
\newblock \href {http://arxiv.org/abs/1102.1957} {\path{arXiv:1102.1957}},
  \href {http://dx.doi.org/10.1103/PhysRevC.84.024906}
  {\path{doi:10.1103/PhysRevC.84.024906}}.

\bibitem{Aad:2014bxa}
{ATLAS Collaboration}, {Measurements of the Nuclear Modification Factor for
  Jets in Pb+Pb Collisions at $\sqrt{s_{\mathrm{NN}}}=2.76$ TeV with the ATLAS
  Detector}, Phys.Rev.Lett. 114~(7) (2015) 072302.
\newblock \href {http://arxiv.org/abs/1411.2357} {\path{arXiv:1411.2357}},
  \href {http://dx.doi.org/10.1103/PhysRevLett.114.072302}
  {\path{doi:10.1103/PhysRevLett.114.072302}}.

\bibitem{Bodwin:1994jh}
G.~T. Bodwin, E.~Braaten, G.~P. Lepage, {Rigorous QCD analysis of inclusive
  annihilation and production of heavy quarkonium}, Phys. Rev. D51 (1995)
  1125--1171, [Erratum: Phys. Rev.D55,5853(1997)].
\newblock \href {http://arxiv.org/abs/hep-ph/9407339}
  {\path{arXiv:hep-ph/9407339}}, \href
  {http://dx.doi.org/10.1103/PhysRevD.55.5853, 10.1103/PhysRevD.51.1125}
  {\path{doi:10.1103/PhysRevD.55.5853, 10.1103/PhysRevD.51.1125}}.

\bibitem{Andronic:2015wma}
A.~Andronic, et~al., {Heavy-flavour and quarkonium production in the LHC era:
  from proton–proton to heavy-ion collisions}, Eur. Phys. J. C76~(3) (2016)
  107.
\newblock \href {http://arxiv.org/abs/1506.03981} {\path{arXiv:1506.03981}},
  \href {http://dx.doi.org/10.1140/epjc/s10052-015-3819-5}
  {\path{doi:10.1140/epjc/s10052-015-3819-5}}.

\bibitem{Brambilla:2004wf}
N.~Brambilla, et~al., {Heavy quarkonium physics}\href
  {http://arxiv.org/abs/hep-ph/0412158} {\path{arXiv:hep-ph/0412158}}.

\bibitem{Brambilla:2008cx}
N.~Brambilla, J.~Ghiglieri, A.~Vairo, P.~Petreczky, {Static quark-antiquark
  pairs at finite temperature}, Phys. Rev. D78 (2008) 014017.
\newblock \href {http://arxiv.org/abs/0804.0993} {\path{arXiv:0804.0993}},
  \href {http://dx.doi.org/10.1103/PhysRevD.78.014017}
  {\path{doi:10.1103/PhysRevD.78.014017}}.

\bibitem{Brambilla:2010cs}
N.~Brambilla, et~al., {Heavy quarkonium: progress, puzzles, and opportunities},
  Eur. Phys. J. C71 (2011) 1534.
\newblock \href {http://arxiv.org/abs/1010.5827} {\path{arXiv:1010.5827}},
  \href {http://dx.doi.org/10.1140/epjc/s10052-010-1534-9}
  {\path{doi:10.1140/epjc/s10052-010-1534-9}}.

\bibitem{Spousta:2015fca}
M.~Spousta, B.~Cole, {Interpreting single jet measurements in Pb $+$ Pb
  collisions at the LHC}, Eur. Phys. J. C76 (2016) 50.
\newblock \href {http://arxiv.org/abs/1504.05169} {\path{arXiv:1504.05169}},
  \href {http://dx.doi.org/10.1140/epjc/s10052-016-3896-0}
  {\path{doi:10.1140/epjc/s10052-016-3896-0}}.

\bibitem{Sjostrand:2014zea}
T.~Sjostrand, S.~Ask, J.~R. Christiansen, R.~Corke, N.~Desai, P.~Ilten,
  S.~Mrenna, S.~Prestel, C.~O. Rasmussen, P.~Z. Skands, {An Introduction to
  PYTHIA 8.2}, Comput. Phys. Commun. 191 (2015) 159--177.
\newblock \href {http://arxiv.org/abs/1410.3012} {\path{arXiv:1410.3012}},
  \href {http://dx.doi.org/10.1016/j.cpc.2015.01.024}
  {\path{doi:10.1016/j.cpc.2015.01.024}}.

\bibitem{Chatrchyan:2014ava}
{CMS Collaboration}, {Measurement of jet fragmentation in PbPb and pp
  collisions at $\sqrt{s_{NN}}=2.76$ TeV}, Phys.Rev. C90~(2) (2014) 024908.
\newblock \href {http://arxiv.org/abs/1406.0932} {\path{arXiv:1406.0932}},
  \href {http://dx.doi.org/10.1103/PhysRevC.90.024908}
  {\path{doi:10.1103/PhysRevC.90.024908}}.

\bibitem{Aad:2014wha}
{ATLAS Collaboration}, {Measurement of inclusive jet charged-particle
  fragmentation functions in Pb+Pb collisions at $\sqrt{s_{NN}} = 2.76$ TeV
  with the ATLAS detector}, Phys.Lett. B739 (2014) 320--342.
\newblock \href {http://arxiv.org/abs/1406.2979} {\path{arXiv:1406.2979}},
  \href {http://dx.doi.org/10.1016/j.physletb.2014.10.065}
  {\path{doi:10.1016/j.physletb.2014.10.065}}.

\bibitem{CMS:2012aa}
{Study of high-pT charged particle suppression in PbPb compared to $pp$
  collisions at $\sqrt{s_{NN}}=2.76$ TeV}, Eur.~Phys.~J.~ C72 (2012) 1945.
\newblock \href {http://arxiv.org/abs/1202.2554} {\path{arXiv:1202.2554}},
  \href {http://dx.doi.org/10.1140/epjc/s10052-012-1945-x}
  {\path{doi:10.1140/epjc/s10052-012-1945-x}}.

\bibitem{MehtarTani:2010ma}
Y.~Mehtar-Tani, C.~A. Salgado, K.~Tywoniuk, {Anti-angular ordering of gluon
  radiation in QCD media}, Phys. Rev. Lett. 106 (2011) 122002.
\newblock \href {http://arxiv.org/abs/1009.2965} {\path{arXiv:1009.2965}},
  \href {http://dx.doi.org/10.1103/PhysRevLett.106.122002}
  {\path{doi:10.1103/PhysRevLett.106.122002}}.

\bibitem{MehtarTani:2011tz}
Y.~Mehtar-Tani, C.~A. Salgado, K.~Tywoniuk, {Jets in QCD Media: From Color
  Coherence to Decoherence}, Phys. Lett. B707 (2012) 156--159.
\newblock \href {http://arxiv.org/abs/1102.4317} {\path{arXiv:1102.4317}},
  \href {http://dx.doi.org/10.1016/j.physletb.2011.12.042}
  {\path{doi:10.1016/j.physletb.2011.12.042}}.

\bibitem{CasalderreySolana:2012ef}
J.~Casalderrey-Solana, Y.~Mehtar-Tani, C.~A. Salgado, K.~Tywoniuk, {New picture
  of jet quenching dictated by color coherence}, Phys.~Lett.~ B725 (2013)
  357--360.
\newblock \href {http://arxiv.org/abs/1210.7765} {\path{arXiv:1210.7765}},
  \href {http://dx.doi.org/10.1016/j.physletb.2013.07.046}
  {\path{doi:10.1016/j.physletb.2013.07.046}}.

\bibitem{Blaizot:2013hx}
J.-P. Blaizot, E.~Iancu, Y.~Mehtar-Tani, {Medium-induced QCD cascade:
  democratic branching and wave turbulence}, Phys.~Rev.~Lett. 111 (2013)
  052001.
\newblock \href {http://arxiv.org/abs/1301.6102} {\path{arXiv:1301.6102}},
  \href {http://dx.doi.org/10.1103/PhysRevLett.111.052001}
  {\path{doi:10.1103/PhysRevLett.111.052001}}.

\bibitem{Konishi:1979cb}
K.~Konishi, A.~Ukawa, G.~Veneziano, {Jet Calculus: A Simple Algorithm for
  Resolving QCD Jets}, Nucl. Phys. B157 (1979) 45--107.
\newblock \href {http://dx.doi.org/10.1016/0550-3213(79)90053-1}
  {\path{doi:10.1016/0550-3213(79)90053-1}}.

\bibitem{Khoze:1997zq}
V.~A. Khoze, S.~Lupia, W.~Ochs, {Perturbative universality in soft particle
  production}, Eur. Phys. J. C5 (1998) 77--90.
\newblock \href {http://arxiv.org/abs/hep-ph/9711392}
  {\path{arXiv:hep-ph/9711392}}, \href
  {http://dx.doi.org/10.1007/s100529800818, 10.1007/s100520050249}
  {\path{doi:10.1007/s100529800818, 10.1007/s100520050249}}.

\bibitem{Capella:1999ms}
A.~Capella, I.~M. Dremin, J.~W. Gary, V.~A. Nechitailo, J.~Tran Thanh~Van,
  {Evolution of average multiplicities of quark and gluon jets}, Phys. Rev. D61
  (2000) 074009.
\newblock \href {http://arxiv.org/abs/hep-ph/9910226}
  {\path{arXiv:hep-ph/9910226}}, \href
  {http://dx.doi.org/10.1103/PhysRevD.61.074009}
  {\path{doi:10.1103/PhysRevD.61.074009}}.

\bibitem{Acosta:2004js}
D.~Acosta, et~al., {Measurement of charged particle multiplicities in gluon and
  quark jets in $p\bar{p}$ collisions at $\sqrt{s} = 1.8$ TeV}, Phys. Rev.
  Lett. 94 (2005) 171802.
\newblock \href {http://dx.doi.org/10.1103/PhysRevLett.94.171802}
  {\path{doi:10.1103/PhysRevLett.94.171802}}.

\bibitem{Abbiendi:1999pi}
G.~Abbiendi, et~al., {Experimental properties of gluon and quark jets from a
  point source}, Eur. Phys. J. C11 (1999) 217--238.
\newblock \href {http://arxiv.org/abs/hep-ex/9903027}
  {\path{arXiv:hep-ex/9903027}}, \href
  {http://dx.doi.org/10.1007/s100520050628} {\path{doi:10.1007/s100520050628}}.

\bibitem{Alioli:2010xd}
S.~Alioli, P.~Nason, C.~Oleari, E.~Re, {A general framework for implementing
  NLO calculations in shower Monte Carlo programs: the POWHEG BOX}, JHEP 06
  (2010) 043.
\newblock \href {http://arxiv.org/abs/1002.2581} {\path{arXiv:1002.2581}},
  \href {http://dx.doi.org/10.1007/JHEP06(2010)043}
  {\path{doi:10.1007/JHEP06(2010)043}}.

\bibitem{Alioli:2010xa}
S.~Alioli, K.~Hamilton, P.~Nason, C.~Oleari, E.~Re, {Jet pair production in
  POWHEG}, JHEP 04 (2011) 081.
\newblock \href {http://arxiv.org/abs/1012.3380} {\path{arXiv:1012.3380}},
  \href {http://dx.doi.org/10.1007/JHEP04(2011)081}
  {\path{doi:10.1007/JHEP04(2011)081}}.

\bibitem{Lai:2010vv}
H.-L. Lai, M.~Guzzi, J.~Huston, Z.~Li, P.~M. Nadolsky, J.~Pumplin, C.~P. Yuan,
  {New parton distributions for collider physics}, Phys. Rev. D82 (2010)
  074024.
\newblock \href {http://arxiv.org/abs/1007.2241} {\path{arXiv:1007.2241}},
  \href {http://dx.doi.org/10.1103/PhysRevD.82.074024}
  {\path{doi:10.1103/PhysRevD.82.074024}}.

\bibitem{Martin:2009iq}
A.~D. Martin, W.~J. Stirling, R.~S. Thorne, G.~Watt, {Parton distributions for
  the LHC}, Eur. Phys. J. C63 (2009) 189--285.
\newblock \href {http://arxiv.org/abs/0901.0002} {\path{arXiv:0901.0002}},
  \href {http://dx.doi.org/10.1140/epjc/s10052-009-1072-5}
  {\path{doi:10.1140/epjc/s10052-009-1072-5}}.

\bibitem{Ball:2010de}
R.~D. Ball, L.~Del~Debbio, S.~Forte, A.~Guffanti, J.~I. Latorre, J.~Rojo,
  M.~Ubiali, {A first unbiased global NLO determination of parton distributions
  and their uncertainties}, Nucl. Phys. B838 (2010) 136--206.
\newblock \href {http://arxiv.org/abs/1002.4407} {\path{arXiv:1002.4407}},
  \href {http://dx.doi.org/10.1016/j.nuclphysb.2010.05.008}
  {\path{doi:10.1016/j.nuclphysb.2010.05.008}}.

\bibitem{Chatrchyan:2011ab}
{CMS Collaboration}, {Measurement of the Inclusive Jet Cross Section in $pp$
  Collisions at $\sqrt{s}=7$ TeV}, Phys.Rev.Lett. 107 (2011) 132001.
\newblock \href {http://arxiv.org/abs/1106.0208} {\path{arXiv:1106.0208}},
  \href {http://dx.doi.org/10.1103/PhysRevLett.107.132001}
  {\path{doi:10.1103/PhysRevLett.107.132001}}.

\bibitem{Chatrchyan:2012np}
{CMS Collaboration}, {Suppression of non-prompt $J/\psi$, prompt $J/\psi$, and
  Y(1S) in PbPb collisions at $\sqrt{s_{NN}}=2.76$ TeV}, JHEP 05 (2012) 063.
\newblock \href {http://arxiv.org/abs/1201.5069} {\path{arXiv:1201.5069}},
  \href {http://dx.doi.org/10.1007/JHEP05(2012)063}
  {\path{doi:10.1007/JHEP05(2012)063}}.

\bibitem{CMS:2012vxa}
{CMS Collaboration}, {J/psi results from CMS in PbPb collisions, with 150mub-1
  data}, {CMS-PAS-HIN-12-014}.

\bibitem{Aad:2010aa}
{ATLAS Collaboration}, {Measurement of the centrality dependence of $J/\psi$
  yields and observation of Z production in lead-lead collisions with the ATLAS
  detector at the LHC}, Phys. Lett. B697 (2011) 294--312.
\newblock \href {http://arxiv.org/abs/1012.5419} {\path{arXiv:1012.5419}},
  \href {http://dx.doi.org/10.1016/j.physletb.2011.02.006}
  {\path{doi:10.1016/j.physletb.2011.02.006}}.

\bibitem{Abelev:2013ila}
{ALICE Collaboration}, {Centrality, rapidity and transverse momentum dependence
  of $J/\psi$ suppression in Pb-Pb collisions at $\sqrt{s_{\rm NN}}$=2.76 TeV},
  Phys. Lett. B734 (2014) 314--327.
\newblock \href {http://arxiv.org/abs/1311.0214} {\path{arXiv:1311.0214}},
  \href {http://dx.doi.org/10.1016/j.physletb.2014.05.064}
  {\path{doi:10.1016/j.physletb.2014.05.064}}.

\bibitem{Khachatryan:2014bva}
{CMS Collaboration}, {Measurement of Prompt $\psi(2S) \to J/\psi$ Yield Ratios
  in Pb-Pb and $p-p$ Collisions at $\sqrt {s_{NN}}=$ 2.76  TeV}, Phys. Rev.
  Lett. 113~(26) (2014) 262301.
\newblock \href {http://arxiv.org/abs/1410.1804} {\path{arXiv:1410.1804}},
  \href {http://dx.doi.org/10.1103/PhysRevLett.113.262301}
  {\path{doi:10.1103/PhysRevLett.113.262301}}.

\bibitem{Adam:2015rba}
{ALICE Collaboration}, {Inclusive, prompt and non-prompt J/$\psi$ production at
  mid-rapidity in Pb-Pb collisions at $\sqrt{s_\mathrm{NN}}$ = 2.76 TeV}, JHEP
  07 (2015) 051.
\newblock \href {http://arxiv.org/abs/1504.07151} {\path{arXiv:1504.07151}},
  \href {http://dx.doi.org/10.1007/JHEP07(2015)051}
  {\path{doi:10.1007/JHEP07(2015)051}}.

\bibitem{ATLAS:2015pua}
{ATLAS Collaboration}, {Study of $J/\psi$ and $\psi(\mathrm{2S})$ production in
  $\sqrt{s_\mathrm{NN}} = 5.02\mathrm{~TeV}$ $p+\rm{Pb}$ and $\sqrt{s} =
  2.76\mathrm{~TeV}$ $pp$ collisions with the ATLAS detector},
  {ATLAS-CONF-2015-023}.

\bibitem{Abelev:2012rv}
{ALICE Collaboration}, {$J/\psi$ suppression at forward rapidity in Pb-Pb
  collisions at $\sqrt{s_{NN}}=2.76$ TeV}, Phys. Rev. Lett. 109 (2012) 072301.
\newblock \href {http://arxiv.org/abs/1202.1383} {\path{arXiv:1202.1383}},
  \href {http://dx.doi.org/10.1103/PhysRevLett.109.072301}
  {\path{doi:10.1103/PhysRevLett.109.072301}}.

\bibitem{Faccioli:2008ir}
P.~Faccioli, C.~Lourenco, J.~Seixas, H.~K. Woehri, {Study of $\psi^\prime$ and
  $\chi_c$ decays as feed-down sources of $J/\psi$ hadro-production}, JHEP 10
  (2008) 004.
\newblock \href {http://arxiv.org/abs/0809.2153} {\path{arXiv:0809.2153}},
  \href {http://dx.doi.org/10.1088/1126-6708/2008/10/004}
  {\path{doi:10.1088/1126-6708/2008/10/004}}.

\bibitem{LHCb:2012af}
R.~Aaij, et~al., {Measurement of the ratio of prompt $\chi_{c}$ to $J/\psi$
  production in $pp$ collisions at $\sqrt{s}=7$ TeV}, Phys. Lett. B718 (2012)
  431--440.
\newblock \href {http://arxiv.org/abs/1204.1462} {\path{arXiv:1204.1462}},
  \href {http://dx.doi.org/10.1016/j.physletb.2012.10.068}
  {\path{doi:10.1016/j.physletb.2012.10.068}}.

\bibitem{ATLAS:2014ala}
{ATLAS Collaboration}, {Measurement of $\chi_{c1}$ and $\chi_{c2}$ production
  with $\sqrt{s}$ = 7 TeV $pp$ collisions at ATLAS}, JHEP 07 (2014) 154.
\newblock \href {http://arxiv.org/abs/1404.7035} {\path{arXiv:1404.7035}},
  \href {http://dx.doi.org/10.1007/JHEP07(2014)154}
  {\path{doi:10.1007/JHEP07(2014)154}}.

\bibitem{Adare:2013ezl}
A.~Adare, et~al., {Nuclear Modification of $\psi', \chi_c$, and $J/\psi$
  Production in d+Au Collisions at $\sqrt{s_{NN}}$=200  GeV}, Phys. Rev.
  Lett. 111~(20) (2013) 202301.
\newblock \href {http://arxiv.org/abs/1305.5516} {\path{arXiv:1305.5516}},
  \href {http://dx.doi.org/10.1103/PhysRevLett.111.202301}
  {\path{doi:10.1103/PhysRevLett.111.202301}}.

\bibitem{Adam:2015iga}
{ALICE Collaboration}, {Rapidity and transverse-momentum dependence of the
  inclusive J/$\psi$ nuclear modification factor in p-Pb collisions at $
  \sqrt{s_{N\ N}} =$ 5.02 TeV}, JHEP 06 (2015) 055.
\newblock \href {http://arxiv.org/abs/1503.07179} {\path{arXiv:1503.07179}},
  \href {http://dx.doi.org/10.1007/JHEP06(2015)055}
  {\path{doi:10.1007/JHEP06(2015)055}}.

\bibitem{Abelev:2013yxa}
{ALICE Collaboration}, {$J/\psi$ production and nuclear effects in p-Pb
  collisions at $\sqrt{S_{NN}}$ = 5.02 TeV}, JHEP 02 (2014) 073.
\newblock \href {http://arxiv.org/abs/1308.6726} {\path{arXiv:1308.6726}},
  \href {http://dx.doi.org/10.1007/JHEP02(2014)073}
  {\path{doi:10.1007/JHEP02(2014)073}}.

\bibitem{Aaij:2013zxa}
R.~Aaij, et~al., {Study of $J/\psi$ production and cold nuclear matter effects
  in $pPb$ collisions at $\sqrt{s_{NN}} = 5$ TeV}, JHEP 02 (2014) 072.
\newblock \href {http://arxiv.org/abs/1308.6729} {\path{arXiv:1308.6729}},
  \href {http://dx.doi.org/10.1007/JHEP02(2014)072}
  {\path{doi:10.1007/JHEP02(2014)072}}.

\bibitem{Eskola:1998df}
K.~J. Eskola, V.~J. Kolhinen, C.~A. Salgado, {The Scale dependent nuclear
  effects in parton distributions for practical applications}, Eur. Phys. J. C9
  (1999) 61--68.
\newblock \href {http://arxiv.org/abs/hep-ph/9807297}
  {\path{arXiv:hep-ph/9807297}}, \href
  {http://dx.doi.org/10.1007/s100520050513, 10.1007/s100529900005}
  {\path{doi:10.1007/s100520050513, 10.1007/s100529900005}}.

\bibitem{Abelev:2012kr}
{ALICE Collaboration}, {Inclusive $J/\psi$ production in $pp$ collisions at
  $\sqrt{s} = 2.76$ TeV}, Phys. Lett. B718 (2012) 295--306, [Erratum: Phys.
  Lett.B748,472(2015)].
\newblock \href {http://arxiv.org/abs/1203.3641} {\path{arXiv:1203.3641}},
  \href {http://dx.doi.org/10.1016/j.physletb.2012.10.078,
  10.1016/j.physletb.2015.06.058} {\path{doi:10.1016/j.physletb.2012.10.078,
  10.1016/j.physletb.2015.06.058}}.

\bibitem{Book:2014mia}
J.~Book, {$J/\psi$ production in Pb–Pb collisions at $\sqrt {s_{NN}}=2.76$
  TeV}, Nucl. Phys. A931 (2014) 591--595.
\newblock \href {http://dx.doi.org/10.1016/j.nuclphysa.2014.09.082}
  {\path{doi:10.1016/j.nuclphysa.2014.09.082}}.

\bibitem{Zhao:2011cv}
X.~Zhao, R.~Rapp, {Medium Modifications and Production of Charmonia at LHC},
  Nucl. Phys. A859 (2011) 114--125.
\newblock \href {http://arxiv.org/abs/1102.2194} {\path{arXiv:1102.2194}},
  \href {http://dx.doi.org/10.1016/j.nuclphysa.2011.05.001}
  {\path{doi:10.1016/j.nuclphysa.2011.05.001}}.

\bibitem{Liu:2009nb}
Y.-p. Liu, Z.~Qu, N.~Xu, P.-f. Zhuang, {J/psi Transverse Momentum Distribution
  in High Energy Nuclear Collisions at RHIC}, Phys. Lett. B678 (2009) 72--76.
\newblock \href {http://arxiv.org/abs/0901.2757} {\path{arXiv:0901.2757}},
  \href {http://dx.doi.org/10.1016/j.physletb.2009.06.006}
  {\path{doi:10.1016/j.physletb.2009.06.006}}.

\bibitem{Ferreiro:2012rq}
E.~G. Ferreiro, {Charmonium dissociation and recombination at LHC: Revisiting
  comovers}, Phys. Lett. B731 (2014) 57--63.
\newblock \href {http://arxiv.org/abs/1210.3209} {\path{arXiv:1210.3209}},
  \href {http://dx.doi.org/10.1016/j.physletb.2014.02.011}
  {\path{doi:10.1016/j.physletb.2014.02.011}}.

\bibitem{Andronic:2011yq}
A.~Andronic, P.~Braun-Munzinger, K.~Redlich, J.~Stachel, {The thermal model on
  the verge of the ultimate test: particle production in Pb-Pb collisions at
  the LHC}, J. Phys. G38 (2011) 124081.
\newblock \href {http://arxiv.org/abs/1106.6321} {\path{arXiv:1106.6321}},
  \href {http://dx.doi.org/10.1088/0954-3899/38/12/124081}
  {\path{doi:10.1088/0954-3899/38/12/124081}}.

\bibitem{Catani:2000ef}
S.~Catani, S.~Dittmaier, Z.~Trocsanyi, {One loop singular behavior of QCD and
  SUSY QCD amplitudes with massive partons}, Phys. Lett. B500 (2001) 149--160.
\newblock \href {http://arxiv.org/abs/hep-ph/0011222}
  {\path{arXiv:hep-ph/0011222}}, \href
  {http://dx.doi.org/10.1016/S0370-2693(01)00065-X}
  {\path{doi:10.1016/S0370-2693(01)00065-X}}.

\bibitem{Dokshitzer:2001zm}
Y.~L. Dokshitzer, D.~E. Kharzeev, {Heavy quark colorimetry of QCD matter},
  Phys. Lett. B519 (2001) 199--206.
\newblock \href {http://arxiv.org/abs/hep-ph/0106202}
  {\path{arXiv:hep-ph/0106202}}, \href
  {http://dx.doi.org/10.1016/S0370-2693(01)01130-3}
  {\path{doi:10.1016/S0370-2693(01)01130-3}}.

\bibitem{Armesto:2003jh}
N.~Armesto, C.~A. Salgado, U.~A. Wiedemann, {Medium induced gluon radiation off
  massive quarks fills the dead cone}, Phys. Rev. D69 (2004) 114003.
\newblock \href {http://arxiv.org/abs/hep-ph/0312106}
  {\path{arXiv:hep-ph/0312106}}, \href
  {http://dx.doi.org/10.1103/PhysRevD.69.114003}
  {\path{doi:10.1103/PhysRevD.69.114003}}.

\bibitem{Baranov:2015laa}
S.~P. Baranov, A.~V. Lipatov, N.~P. Zotov, {Prompt charmonia production and
  polarization at LHC in the NRQCD with $k_T$ -factorization. Part I: $\psi
  (2S)$ meson}, Eur. Phys. J. C75~(9) (2015) 455.
\newblock \href {http://arxiv.org/abs/1508.05480} {\path{arXiv:1508.05480}},
  \href {http://dx.doi.org/10.1140/epjc/s10052-015-3689-x}
  {\path{doi:10.1140/epjc/s10052-015-3689-x}}.

\end{thebibliography}
\bibliographystyle{elsarticle-num}
\end{document}